\documentclass[12pt]{ronbun}

\usepackage{amsmath,amssymb,graphicx,color}
\usepackage{cite}
\usepackage{bm}
\usepackage{dcolumn}

\newcommand{\nn}{\nonumber}
\newcommand{\e}{{\rm e}}

\newcommand{\del}{\delta}

\newcommand{\al}{\alpha}

\numberwithin{equation}{section}

\begin{document}

\begin{flushright}
\parbox{4.2cm}
{KEK-TH-930 \hfill \\
{\tt hep-th/0401014}
 }
\end{flushright}

\vspace*{1.1cm}

\begin{center}
 \Large\bf Thermodynamics of Fuzzy Spheres \\
in PP-wave Matrix Model 
\end{center}
\vspace*{1.5cm}
\centerline{\large  Hyeonjoon Shin$^{\dagger a}$ and Kentaroh
Yoshida$^{\ast b}$}
\begin{center}
$^{\dagger}$\emph{ BK 21 Physics Research
    Division and Institute of Basic Science,\\ 
    Sungkyunkwan University,
    Suwon 440-746, South Korea
}\\ 
$^{\ast}$\emph{ Theory Division, High Energy Accelerator Research 
Organization (KEK),\\
Tsukuba, Ibaraki 305-0801, Japan.} 
\\
\vspace*{1cm}
 $^{a}$hshin@newton.skku.ac.kr \qquad $^{b}$kyoshida@post.kek.jp
\end{center}

\vspace*{1.5cm}

\centerline{\bf Abstract} \vspace*{0.5cm} 
We discuss thermodynamics of fuzzy spheres in a matrix model on a
pp-wave background. The exact free energy in the fuzzy sphere vacuum
is computed in the $\mu\rightarrow\infty$ limit for an arbitrary
matrix size $N$.  The trivial vacuum dominates the fuzzy sphere vacuum
at low temperature while the fuzzy sphere vacuum is more stable than
the trivial vacuum at sufficiently high temperature. Our result
supports that the fluctuations around the trivial vacuum would
condense to form an irreducible fuzzy sphere above a certain
temperature.

\vfill

\thispagestyle{empty}
\setcounter{page}{0}

\newpage 

\section{Introduction}

In a recent study of string theories and M-theory, a matrix model on a
pp-wave background was proposed by Berenstein-Maldacena-Nastase
\cite{BMN}, and it has been intensively studied. The background of
this matrix model is given by the maximally supersymmetric pp-wave
background \cite{KG}:
\begin{eqnarray}
\label{pp}
ds^2 &=& -2dx^+dx^- -\left(
\sum_{i=1}\left(\frac{\mu}{3}\right)^2 (x^i)^2 + \sum_{a=4}^6\left(
\frac{\mu}{6}\right)^2 (x^a)^2
\right)(dx^+)^2 + \sum_{I=1}^9(dx^{I})^2\,, \\
F_{+123} &=& \mu\,. \nn
\end{eqnarray}
The matrix model on this background has a supersymmetric fuzzy sphere
solution (which is called ``giant graviton'') due to the Myers effects
\cite{Myers}, because the constant 4-form flux is equipped with.  The
classical energy of this solution is zero and hence 
the fuzzy sphere can appear in classical vacua without loss of energy. 
Namely, 
the classical vacua of the pp-wave matrix model are enriched with fuzzy
spheres, and it may be interesting to look deeper into the dynamics of
fuzzy sphere solution.

In fact, we have studied the quantum stability of fuzzy sphere (giant
graviton) solution in the pp-wave matrix model by using the path
integral formulation \cite{SY3,HSKY}. In the work \cite{SY3} the
quantum stability of a supersymmetric fuzzy sphere and instability of
non-supersymmetric fuzzy sphere in the $2\times 2$ matrix case. Then,
we considered the interaction between two fuzzy spheres \cite{HSKY} in
the limit $\mu\rightarrow\infty$ \cite{DSR}.  In part, this work was
the generalization of the result of \cite{SY3} to an arbitrary matrix
size $N$ case.

In this paper, we study the thermodynamics of fuzzy spheres by using
the method formulated in the work \cite{HSKY}. We present the exact
free energy in the fuzzy sphere vacuum by using the limit
$\mu\rightarrow\infty$ for an arbitrary matrix size $N$\,. It is found
from numerical plots of the free energy that the fuzzy sphere vacuum
is more stable than trivial vacuum at sufficiently high temperature
while the trivial vacuum dominates the fuzzy sphere vacuum at low
temperature.  Namely, there would be a critical temperature, above
which the free energy in the fuzzy sphere vacuum is smaller than that
in the trivial one. This temperature depends on a matrix size $N$, and
it grows as the value of $N$ becomes large. In particular, our
approximate evaluation suggests that such a change in vacuum type
would not appear in the $N\rightarrow \infty$ limit. This result is
consistent with the supergravity analysis. It is because the large $N$
limit means that the supergravity description is valid \cite{DSR} but
the fuzzy sphere configuration could not be described in the context
of supergravity.  Furthermore, we discuss that the fuzzy sphere
belonging to an $N$-dimensional irreducible representation of fuzzy
sphere is more stable than the reducible one at sufficiently 
high temperature.

This paper is organized as follows: In section 2, we briefly introduce
a pp-wave matrix model, and explain the method to calculate one-loop
corrections around fuzzy sphere solutions.  In section 3, the exact
free energy is calculated.  Then we numerically evaluate the
difference of free energies around the trivial and fuzzy sphere vacua.
Section 4 is devoted to a conclusion and discussions.

\section{One-loop Calculation in the PP-wave Matrix Model}

We will introduce a pp-wave matrix model \cite{BMN}, which 
is basically composed of two parts: 
\begin{equation}
S_{pp} = S_\mathrm{flat} + S_\mu ~,
\label{pp-bmn}
\end{equation}
where each part of the action on the right hand side is given by
\begin{align}
S_\mathrm{flat} & = \int dt \mathrm{Tr} 
\left( \frac{1}{2R} D_t X^I D_t X^I + \frac{R}{4} ( [ X^I, X^J] )^2
      + i \Theta^\dagger D_t \Theta 
      - R \Theta^\dagger \gamma^I [ \Theta, X^I ]
\right) ~,
  \notag \\
S_\mu &= \int dt \mathrm{Tr}
\left( 
      -\frac{1}{2R} \left( \frac{\mu}{3} \right)^2 (X^i)^2
      -\frac{1}{2R} \left( \frac{\mu}{6} \right)^2 (X^a)^2
      - i \frac{\mu}{3} \epsilon^{ijk} X^i X^j X^k
      - i \frac{\mu}{4} \Theta^\dagger \gamma^{123} \Theta
\right) ~.
\label{o-action}
\end{align}
Here, $R$ is the radius of circle compactification along $x^-$ and $D_t$
is the covariant derivative with the gauge field $A$,
\begin{equation}
D_t = \partial_t - i [A, \: ] ~.
\end{equation}
It is convenient to
rescale the gauge field and parameters as
\begin{equation}
A \rightarrow R A ~,~~~ 
t \rightarrow \frac{1}{R} t ~,~~~
\mu \rightarrow R \mu ~.
\label{res0}
\end{equation}
In this matrix model, classical equations of motion allow the following
membrane fuzzy sphere or giant graviton solution:
\begin{equation}
X^i_\mathrm{sphere} = \frac{\mu}{3} J^i ~,
\label{fuzzy}
\end{equation}
where $J^i$ satisfies the $SU(2)$ algebra $[ J^i, J^j ] = i
\epsilon^{ijk} J^k$\,.  The reason why this solution is possible is
basically that the matrix field $X^i$ feels an extra force due to the
Myers interaction which may stabilize the oscillatory force.  The
fuzzy sphere solution $X^i_\mathrm{sphere}$ preserves the 16 dynamical
supersymmetries of the pp-wave and hence is 1/2-BPS object.  We note
that actually there is another fuzzy sphere solution of the form
$\frac{\mu}{6} J^i$.  However, it has been shown that such solution
does not have quantum stability and is thus non-BPS object \cite{SY3}.

From now on, we will briefly review the calculus of one-loop quantum
corrections. By the use of the background field method, the pp-wave
matrix model can be expanded around the general bosonic background,
which is supposed to satisfy the classical equations of motion.

We first split the matrix quantities into as follows:
\begin{equation}
X^I = B^I + Y^I ~,~~~ \Theta = F + \Psi ~,
\label{cl+qu}
\end{equation}
where $B^I$ and $F$ are the classical background fields while $Y^I$ and
$\Psi$ are the quantum fluctuations around them.  The fermionic
background $F$ is taken to vanish from now on, since we will only
consider the purely bosonic background. In order to carry out the path
integration for the fields, let us take the background field gauge which
is usually chosen in the matrix model calculation as
\begin{equation}
D_\mu^{\rm bg} A^\mu_{\rm qu} \equiv 
D_t A + i [ B^I, X^I ] = 0 ~.
\label{bg-gauge}
\end{equation}
Then the corresponding gauge-fixing $S_\mathrm{GF}$ and Faddeev-Popov
ghost $S_\mathrm{FP}$ terms are given by
\begin{equation}
S_\mathrm{GF} + S_\mathrm{FP} =  \int\!dt \,{\rm Tr}
  \left(
      -  \frac{1}{2} (D_\mu^{\rm bg} A^\mu_{\rm qu} )^2 
      -  \bar{C} \partial_{t} D_t C + [B^I, \bar{C}] [X^I,\,C]
\right) ~.
\label{gf-fp}
\end{equation}

Now by inserting the decomposition of the matrix fields (\ref{cl+qu})
into the matrix model action, we get the gauge fixed plane-wave action
$S$ $(\equiv S_{pp} + S_\mathrm{GF} + S_\mathrm{FP})$ expanded around
the background.  The resulting acting is read as
\begin{equation}
S =  S_0 + S_2 + S_3 + S_4 ~,
\end{equation}
where $S_n$ represents the action of order $n$ with respect to the
quantum fluctuations and, for each $n$, its expression is
\begin{align}
S_0 = \int dt \, \mathrm{Tr} \bigg[ \,
&      \frac{1}{2}(\dot{B}^I)^2  
        - \frac{1}{2} \left(\frac{\mu}{3}\right)^2 (B^i)^2 
        - \frac{1}{2} \left(\frac{\mu}{6}\right)^2 (B^a)^2 
        + \frac{1}{4}([B^I,\,B^J])^2
        - i \frac{\mu}{3} \epsilon^{ijk} B^i B^j B^k 
    \bigg] ~,
\notag \\
S_2 = \int dt \, \mathrm{Tr} \bigg[ \,
&       \frac{1}{2} ( \dot{Y}^I)^2 - 2i \dot{B}^I [A, \, Y^I] 
        + \frac{1}{2}([B^I , \, Y^J])^2 
        + [B^I , \, B^J] [Y^I , \, Y^J]
        - i \mu \epsilon^{ijk} B^i Y^j Y^k
\notag \\
&       - \frac{1}{2} \left( \frac{\mu}{3} \right)^2 (Y^i)^2 
        - \frac{1}{2} \left( \frac{\mu}{6} \right)^2 (Y^a)^2 
        + i \Psi^\dagger \dot{\Psi} 
        -  \Psi^\dagger \gamma^I [ \Psi , \, B^I ] 
        -i \frac{\mu}{4} \Psi^\dagger \gamma^{123} \Psi  
\notag \\ 
&       - \frac{1}{2} \dot{A}^2  - \frac{1}{2} ( [B^I , \, A])^2 
        + \dot{\bar{C}} \dot{C} 
        + [B^I , \, \bar{C} ] [ B^I ,\, C] \,
     \bigg] ~,
\notag \\
S_3 = \int dt \, \mathrm{Tr} \bigg[
&       - i\dot{Y}^I [ A , \, Y^I ] - [A , \, B^I] [ A, \, Y^I] 
        + [ B^I , \, Y^J] [Y^I , \, Y^J] 
        +  \Psi^\dagger [A , \, \Psi] 
\notag \\
&       -  \Psi^\dagger \gamma^I [ \Psi , \, Y^I ] 
        - i \frac{\mu}{3} \epsilon^{ijk} Y^i Y^j Y^k
        - i \dot{\bar{C}} [A , \, C] 
        +  [B^I,\, \bar{C} ] [Y^I,\,C]  \,
     \bigg] ~,
\notag \\
S_4 = \int dt \, \mathrm{Tr} \bigg[
&       - \frac{1}{2} ([A,\,Y^I])^2 + \frac{1}{4} ([Y^I,\,Y^J])^2 
     \bigg] ~.
\label{bgaction} 
\end{align}

For the justification of one-loop computation or the semi-classical
analysis, it should be made clear that $S_3$ and $S_4$ of
Eq.~(\ref{bgaction}) can be regarded as perturbations.  For this
purpose, following \cite{DSR}, we rescale the fluctuations and
parameters as
\begin{gather}
A   \rightarrow \mu^{-1/2} A   ~,~~~
Y^I \rightarrow \mu^{-1/2} Y^I ~,~~~
C  \rightarrow \mu^{-1/2} C   ~,~~~
\bar{C} \rightarrow \mu^{-1/2} \bar{C} ~,~~~ 
t \rightarrow \mu^{-1} t ~.
\label{rescale}
\end{gather}
Under this rescaling, the action $S$ in the fuzzy sphere background becomes
\begin{align}
S =  S_2 + \mu^{-3/2} S_3 + \mu^{-3} S_4 ~,
\label{ssss}
\end{align}
where $S_2$, $S_3$ and $S_4$ do not have $\mu$ dependence.  Now it is
obvious that, in the large $\mu$ limit, $S_3$ and $S_4$ can be treated
as perturbations and the one-loop computation gives the sensible
result. Note that the analysis in the $S_2$ part is exact in the
$\mu\rightarrow \infty$ limit.

We can calculate the exact spectra around an $N$-dimensional
irreducible fuzzy sphere in the $\mu \rightarrow \infty$ limit, by
following the method in the work \cite{DSR} (For the detailed
calculation, see \cite{DSR,HSKY}).  The spectra are summarized in
Tabs.\,\ref{boson:tab} and \ref{fermion:tab}.

\begin{table}
 \begin{center}
{
\small\bf 
  \begin{tabular}{|c|c|c|c|}
\hline 
    {\boldmath $SO(3)$} Bosons & Mass & Degeneracy & Spin \\
\hline 
$u_{jm}$ & $\frac{1}{3}(j+1)$ & $2j+1$ & $0\leq j \leq N -2$ \\
$v_{jm}$ & $\frac{1}{3}j$ & $2j+1$ & $ 1\leq j \leq N $ \\
$w_{jm}$ & $\frac{1}{3}\sqrt{j(j+1)}$ & $2j+1$ & $1\leq j \leq N -1$
   \\
\hline \hline
{\boldmath $SO(6)$} Bosons & Mass & Degeneracy & Spin \\
\hline
$z^a_{jm}~~(a=1,\dots,6)$ & $\frac{1}{3}\left(j + \frac{1}{2}\right)$
   & $6(2j+1)$ & $0\leq j \leq N -1$ \\
\hline
\hline 
Gauge Field & Mass & Degeneracy & Spin \\
\hline
$z^0_{jm}$ & $\frac{1}{3}\sqrt{j(j+1)}$ & $2j+1$ & $0\leq j \leq N -1$ \\ 
\hline
\end{tabular}
}
\caption{Bosonic spectrum around an irreducible fuzzy sphere}
\label{boson:tab}
\end{center}
\begin{center}
{
\small\bf
\begin{tabular}{|c|c|c|c|}
\hline
Fermions & Mass & Degeneracy & Spin \\
\hline 
$\pi_{jm}$ & $\frac{1}{3}\left(j + \frac{3}{4}\right)$ & $8(2j+1)$ & 
$ \frac{1}{2}\leq j \leq N - \frac{3}{2}$ \\  
$\eta_{jm}$ & $\frac{1}{3}\left(j+\frac{1}{4}\right)$ &
   $8(2j+1)$ & $\frac{1}{2}\leq j \leq N - \frac{1}{2}$ \\
\hline\hline 
Ghosts & Mass & Degeneracy & Spin \\
\hline 
$C_{jm}~(\bar{C}_{jm})$ & $\frac{1}{3}\sqrt{j(j+1)}$ 
& $2j+1$ & $0 \leq j \leq N - 1$ \\
\hline
  \end{tabular}
}
\caption{Fermionic spectrum around an irreducible fuzzy sphere}
\label{fermion:tab}
 \end{center}
\end{table}

\section{Thermal Correction and Free Energy}

We now calculate the one-loop correction in the case that the system
couples to a thermal bath. In the zero temperature case, a
supersymmetric fuzzy sphere is quantum mechanically stable at one-loop
level as shown in \cite{SY3,HSKY}. In particular, the quantum
corrections are just canceled out due to supersymmetry.  All the
supersymmetries are broken down when we consider the finite
temperature case, and the fuzzy sphere is no more protected by
supersymmetries against quantum fluctuations. Note, however, that
supersymmetry breaking does not necessary imply the instability of
fuzzy sphere because the word ``instability'' means the presence of
negative eigen-modes around the classical configuration.  In the work
\cite{Huang}, Huang calculated the free energy by using the $2\times
2$ matrix formulation introduced in \cite{SY3}.  From now on, we
utilize a more general formulation proposed in \cite{HSKY} that is
based on the matrix perturbation theory in \cite{DSR}, and calculate
the exact free energy in an arbitrary matrix size $N$ case. We can
find out some advantages and new physics in our general formulation,
as we will see below.

In order to consider the thermal system with temperature $T$, let us
move to the Euclidean formulation via the Wick rotation $t\rightarrow
it$, and compactify the Euclidean time direction with a periodicity
$\beta \equiv 1/T$.  Note that $T$ is a dimensionless parameter
because of the scaling of time variable $t \rightarrow R^{-1}t$,
(\ref{res0}).  This compactification leads us to encounter the
following summation, instead of the momentum integral,
\begin{eqnarray}
\sum_{n}\ln\left[\left(\frac{2\pi n}{\beta}\right)^2 + M^2\right]\,,
\end{eqnarray}
where $M$ is a mass parameter, and the index $n$ takes integer for
bosons and half-integer for fermions. We can easily compute this
summation by using the formulae\footnote{In the derivation of this
  formulae we have dropped out some infinite constants containing no
  physical parameters. Note, however, that these divergent constants
  are canceled with each other due to supersymmetries at zero
  temperature, even if we should keep them.}
\begin{eqnarray}
\sum_{n={\rm integer}}\ln\left(n^2\pi^2 + M'{}^2\right) = 2\ln\sinh M' \quad
 (\mbox{for bosons})\,,
 \\
\sum_{n={\rm half~integer}}\ln\left(n^2\pi^2 + M'{}^2\right) = 2\ln\cosh M'
 \quad (\mbox{for fermions})\,,
\end{eqnarray} 
and the fact that the fuzzy sphere configuration under our consideration
is supersymmetric.  The free energy $F = - (1/\beta)\ln Z$ is
represented by
\begin{eqnarray}
F &=& T\sum_{i\in Y,A}\left(\frac{N_i}{2}\right)\left[
2\ln\left(1 - \e^{-\beta M_i}\right) 
\right] \nn \\
&& - T\sum_{i\in C,\psi}\left( N_i \del_{i,C} 
+ \frac{N_i\del_{i,\Psi}}{4} \right)
\left[
2 \ln\left(1 + \e^{-\beta M_i}\right)
\right]\,, \label{general}
\end{eqnarray}
where $M_i$ and $N_i$ are the mass and degeneracy of the $i$ type of
fluctuation field. The symbols $Y$ and $A$ denote the bosonic
fluctuations of $X$ and $A$, respectively. The $\Psi$ and $C$ denote the
fermionic fluctuations of $\Theta$ and $C$\,. The free energy usually
contains the zero temperature part, but this part does not appear in the
present case since the fuzzy sphere background is supersymmetric at zero
temperature.

\subsection{Free energies in irreducible vacua} 

To begin with, we shall consider the free energy in the irreducible
vacuum.  By inserting the values of mass and degeneracy listed in
Tables \ref{boson:tab} and \ref{fermion:tab}, we can obtain the
concrete expression of the free energy as follows:
\begin{eqnarray}
\label{free-fuzzy}
 \beta F &=& \sum_{j=0}^{N -2}(2j+1) \ln\left(1 - 
\e^{- \frac{\beta}{3}(j+1)}\right) 
+ \sum_{j=1}^{N}(2j+1) \ln\left(1 - 
\e^{- \frac{\beta}{3}j}\right) \\
&&  + \sum_{j=1}^{N - 1}(2j+1) \ln\left(1 - 
\e^{- \frac{\beta}{3}\sqrt{j(j+1)}}\right)  
 + \sum_{j=0}^{N - 1}6(2j+1) \ln\left(1 - 
\e^{- \frac{\beta}{3}\left(j + \frac{1}{2}\right)}\right)  \nn \\
& & + \sum_{j=1}^{N - 1}(2j+1) \ln\left(1 - 
\e^{- \frac{\beta}{3}\sqrt{j(j+1)}}\right)
- \sum_{j=\frac{1}{2}}^{N - \frac{3}{2}} 
4(2j+1) \ln\left(1 + \e^{- \frac{\beta}{3}
\left(j + \frac{3}{4}\right)}\right) \nn \\
&& - \sum_{j=\frac{1}{2}}^{N - \frac{1}{2}} 
4(2j+1) \ln\left(1 + \e^{- \frac{\beta}{3}
\left(j + \frac{1}{4}\right)}\right) 
 - \sum_{j=1}^{N - 1} 
2(2j+1) \ln\left(1 + \e^{- \frac{\beta}{3}
\sqrt{j(j+1)}}\right) \,, \nn 
\end{eqnarray}
where we note that the $j=0$ modes in the gauge field and ghost parts
are dropped out, since the massless part does not contribute to the
finite temperature effect.

\begin{figure}
 \begin{center}
  \includegraphics[scale=1.0]{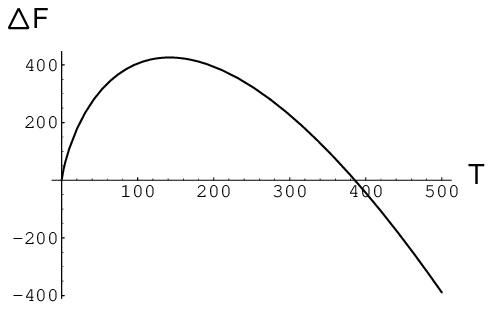} {\footnotesize (i)~$N=2$}  
  \includegraphics[scale=1.0]{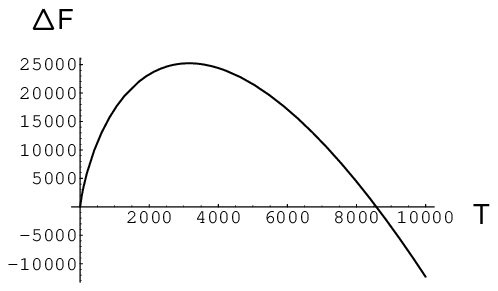} {\footnotesize (ii)~$N=3$} 
  \includegraphics[scale=1.0]{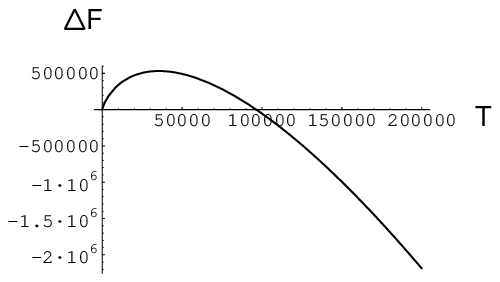} {\footnotesize (iii)~$N=4$} 
  \includegraphics[scale=1.0]{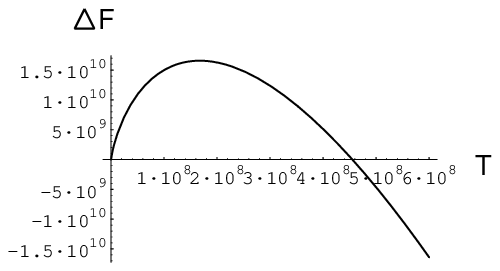} {\footnotesize (iv)~$N=10$} 
\caption{The numerical plots of the free energy difference in the
  $N=2,3,4,10$ cases.} 
\label{plot:fig}
 \end{center}
\end{figure}

On the other hand, the free energy in the trivial vacuum given by $X^I
= 0$ is obtained as
\begin{eqnarray}
\label{free-trivial}
F_{0} &=& 3T N^2\ln\left(1-\e^{-\frac{1}{3T}}\right) + 6T N^2\ln
\left(1-\e^{-\frac{1}{6T}}\right) - 8TN^2\ln\left(
1+\e^{-\frac{1}{4T}}\right)
\,.
\end{eqnarray}
Let us now introduce the difference of free energy difference $\Delta
F \equiv F - F_0$, and evaluate it numerically. We take four cases of
$N=2,3,5,10$. Then the numerical plots for them are as in
Fig.\,\ref{plot:fig} (i)-(iv).
The difference $\Delta F$ has an additional vanishing value apart from
$T =0$. The free energy in the trivial vacuum is dominant in the low
temperature region, and the trivial vacuum is more stable than the
fuzzy sphere one.  The free energy in the fuzzy sphere vacuum is,
however, smaller than that in the trivial vacuum above a certain
temperature\footnote{ We call below this temperature a critical
  temperature, though this usage of ``critical'' is not accurate.}.
We note that our numerical plot in the case of $N=2$ shown in
Fig.\,\ref{plot:fig}~(i) recovers the result of Huang \cite{Huang}.
We see that the critical temperature grows as the matrix size $N$
becomes large.  Although we do not have the exact analytical
estimation for the critical temperature, one may see that the critical
temperature increases quite rapidly as we increase $N$.

It is possible to evaluate asymptotic forms of free energies at
sufficiently high and low temperature.  In the high temperature regime,
the leading terms of free energies in the fuzzy sphere vacuum and
trivial vacuum are, respectively, given by
\begin{eqnarray}
\label{3.7}
&& F \sim -(10 N^2 - 1) T\ln T \quad (\mbox{fuzzy sphere vacuum})\,, \\
&& F_{0} \sim - 9 TN^2 \ln T\quad (\mbox{trivial vacuum})\,.
\label{3.8}
\end{eqnarray}
In the low temperature regime, the leading contributions to free
energies are
\begin{eqnarray}
\label{3.9}
&& F \sim -6 T \e^{-\frac{1}{6T}} \quad (\mbox{fuzzy sphere vacuum})\,,
 \\
&& F_{0} \sim -6 T N^2 \e^{-\frac{1}{6T}} \quad (\mbox{trivial vacuum})\,. 
\label{3.10}
\end{eqnarray}
The asymptotic form (\ref{3.9}) does not depend on the matrix size
$N$\,.  When we put $N=2$ into Eqs.\,(\ref{3.7}), (\ref{3.8}) and
(\ref{3.10}), the results of \cite{Huang} are recovered. From the
above asymptotic forms, we see that what type of vacuum has smaller
free energy at sufficiently low and high temperatures, and may confirm
the argument that there would be a critical temperature.

It is also possible to evaluate roughly the free energy in the large
$N$ limit.  The free energy in the fuzzy sphere vacuum
(\ref{free-fuzzy}) may be approximatively described by the integral
expression
\begin{eqnarray}
F_{N\gg 2} \sim 20 T \int^N_0\!\!dx\,x\ln\left[
\tanh\left(\frac{x}{6T}\right)
\right]\,.
\end{eqnarray}  
The free energy in the trivial vacuum is proportional to $N^2$, and
hence we may adopt (\ref{free-trivial}) as the leading contribution in
the large $N$ limit. When we numerically analyze the difference of
free energies $\Delta F_{N\gg 2} \equiv F_{N\gg 2} - F_{0}$\,, it is
always positive except for $T=0$. That is, the trivial vacuum is more
stable than the fuzzy sphere vacuum for any values of the temperature
when we consider the large $N$ limit. This result supports that the
critical temperature would be infinite in the large $N$ limit, and
hence the supermembrane theory, which is realized as the $N\rightarrow
\infty$ limit of the matrix model, might not show critical behavior in
its thermodynamics (for the relation between supermembrane theory and
matrix model on pp-waves, see \cite{DSR,SY}). On the other hand, the
large $N$ limit corresponds to the region where the supergravity
analysis\footnote{For the supergravity analysis in the
  eleven-dimensional pp-wave background, see the work \cite{KY}. An
  approach from the type IIA plane-wave supergravity analysis
  \cite{KS} is also helpful. }  is valid \cite{DSR}. We emphasize that
our result leads to the consistent picture in the large $N$ limit,
because the critical phenomenon in the thermodynamics of fuzzy spheres
would not be described in the context of supergravity.

\subsection{Free energies in reducible vacua}

\begin{table}
 \begin{center}
{
\small\bf 
  \begin{tabular}{|c|c|c|c|}
\hline 
    {\boldmath $SO(3)$} Bosons & Mass & Degeneracy & Spin \\
\hline 
$\al^{jm}_{kl}$ & $\frac{1}{3}(j+1)$ & $2j+1$ & $ \frac{1}{2}|N_k - N_l|-1 
\leq j \leq \frac{1}{2}(N_k + N_l) -2$ \\
$\beta^{jm}_{kl}$ & $\frac{1}{3}j$ & $2j+1$ & $ \frac{1}{2}|N_k - N_l| +1 
\leq j \leq \frac{1}{2}(N_k + N_l) $ \\
$\omega^{jm}_{kl}$ & $\frac{1}{3}\sqrt{j(j+1)}$ & $2j+1$ & $ \frac{1}{2}|N_k - N_l| 
\leq j \leq \frac{1}{2}(N_k + N_l) -1$
   \\
\hline \hline
{\boldmath $SO(6)$} Bosons & Mass & Degeneracy & Spin \\
\hline
$(x^a_{kl})_{jm}~~(a=1,\dots,6)$ & $\frac{1}{3}\left(j + \frac{1}{2}\right)$
   & $6(2j+1)$ & $ \frac{1}{2}|N_k - N_l|\leq j \leq 
\frac{1}{2}(N_k + N_l)-1$ \\
\hline
\hline 
Gauge Field  & Mass & Degeneracy & Spin \\
\hline
$(\phi^0_{kl})_{jm}$ & $\frac{1}{3}\sqrt{j(j+1)}$ & $2j+1$ 
& $\frac{1}{2}|N_k - N_l| \leq j \leq \frac{1}{2}(N_k + N_l) -1$ \\ 
\hline
\end{tabular}
}
\caption{Bosonic spectrum around reducible fuzzy sphere vacua}
\label{boson:tab2}
\end{center}
\begin{center}
{
\small\bf
\begin{tabular}{|c|c|c|c|}
\hline
Fermions & Mass & Degeneracy & Spin \\
\hline 
$\chi^{A~jm}_{kl}$ & $\frac{1}{3}\left(j + \frac{3}{4}\right)$ & 
$8(2j+1)$ & 
$ \frac{1}{2}|N_k - N_l| -\frac{1}{2} \leq j \leq \frac{1}{2}(N_k + N_l) 
- \frac{3}{2}$ \\  
$\eta^{jm}_{A~kl}$ & $\frac{1}{3}\left(j+\frac{1}{4}\right)$ &
   $8(2j+1)$ & $\frac{1}{2}|N_k - N_l| + \frac{1}{2}
\leq j \leq \frac{1}{2}(N_k +N_l) - \frac{1}{2}$ \\
\hline\hline 
Ghosts & Mass & Degeneracy & Spin \\
\hline 
$C_{jm}~(\bar{C}_{jm})$ & $\frac{1}{3}\sqrt{j(j+1)}$ 
& $2j+1$ & $\frac{1}{2}|N_k - N_l| \leq j \leq 
\frac{1}{2}(N_k + N_l) -1$ \\
\hline
  \end{tabular}
}
\caption{Fermionic spectrum around reducible fuzzy sphere vacua}
\label{fermion:tab2}
 \end{center}
\end{table} 

Let us now include the reducible vacuum in our consideration and
compare the free energies of irreducible and reducible representations
of the fuzzy sphere solution. We can carry out this kind of discussion,
since the formulation with the arbitrary matrix size is now utilized.
Let us consider an $N$-dimensional reducible representation of $SU(2)$
described by a block-diagonal matrix
\begin{eqnarray}
X^i = 
\begin{pmatrix}
J_1^i & &  \\
& \ddots & \\
&&  J^i_K 
\end{pmatrix}
\,,
\end{eqnarray}
where $J_l^i~(i=1,2,3;~l=1,\dots,K)$ are generators of the
$N_l$-dimensional irreducible representation of $SU(2)$ and $N_1 +
\dots + N_{K} = N$\,.  The spectra around reducible vacua are listed
in Tables\,\ref{boson:tab2} and \ref{fermion:tab2}. It should be noted
that they are those for the block off-diagonal fluctuations.

Instead of investigating the general situation, we take the simplest
non-trivial case, that is, $K=2$, which is enough for comparison with
other types of vacua. Then the free energy for the $N=N_1 + N_2~(N_1
\neq N_2)$ case is given by
\begin{eqnarray}
&& \beta F = \sum_{j=\frac{1}{2}|N_1 - N_2|-1}^{\frac{1}{2}(N_1 + N_2) -2}
2(2j+1)\ln\left(1 - \e^{- \frac{\beta}{3}(j+1)}\right) 
+ \sum_{j=\frac{1}{2}|N_1 - N_2| +1}^{\frac{1}{2}(N_1 + N_2)}
2(2j+1) \ln\left(1 - \e^{- \frac{\beta}{3}j}\right)  \nn \\
&&  + \sum_{j=\frac{1}{2}|N_1 - N_2|}^{\frac{1}{2}(N_1 + N_2) -1}
2(2j+1) \ln\left(1 - \e^{- \frac{\beta}{3}\sqrt{j(j+1)}}\right)  
 + \sum_{j=\frac{1}{2}|N_1 - N_2|}^{\frac{1}{2}(N_1 + N_2) -1} 
12(2j+1) \ln\left(1 -  \e^{- \frac{\beta}{3}
\left(j + \frac{1}{2}\right)}\right)  \nn \\
& & + \sum_{j=\frac{1}{2}|N_1 - N_2|}^{\frac{1}{2}(N_1 +N_2) -1}
2(2j+1) \ln\left(1 - \e^{- \frac{\beta}{3}\sqrt{j(j+1)}}\right)
- 2\sum_{j=\frac{1}{2}|N_1 - N_2|-\frac{1}{2}}^{\frac{1}{2}(N_1 + N_2) 
- \frac{3}{2}} 
8(2j+1) \ln\left(1 + \e^{- \frac{\beta}{3}
\left(j + \frac{3}{4}\right)}\right) \nn \\
&& - \sum_{j=\frac{1}{2}|N_1 - N_2| 
+ \frac{1}{2}}^{\frac{1}{2}(N_1 + N_2) - \frac{1}{2}} 
8(2j+1) \ln\left(1 + \e^{- \frac{\beta}{3}
\left(j + \frac{1}{4}\right)}\right) 
- \sum_{j=\frac{1}{2}|N_1 - N_2|}^{\frac{1}{2}(N_1 + N_2) - 1} 
4(2j+1) \ln\left(1 + \e^{- \frac{\beta}{3}
\sqrt{j(j+1)}}\right)  \nn \\
&& + \beta F_{N_1} + \beta F_{N_2}\,. 
\end{eqnarray}
where $F_{N_1}~(F_{N_2})$ is the free
energy of the fuzzy sphere of size $N_1 \times N_1~(N_2 \times N_2)$ given by 
Eq.\,(3.5).

Thus, by using the spectra in Tabs.\,\ref{boson:tab},
\ref{fermion:tab}, \ref{boson:tab2}, \ref{fermion:tab2} and general
formula (\ref{general}), free energies can be computed in general
reducible vacua of arbitrary $N_1$ and $N_2$ with $N_1+N_2=N$.
Although the expressions of free energies are so complicated that we
could not evaluate analytically, it is possible to investigate the
free energy difference numerically between irreducible and reducible
representations. Since it is also difficult to find out the stable
vacuum for general matrix size $N$, we will study the matrix size
$N=3$ and $N=4$ cases, as examples.

\begin{figure}
 \begin{center}
\includegraphics[scale=1.0]{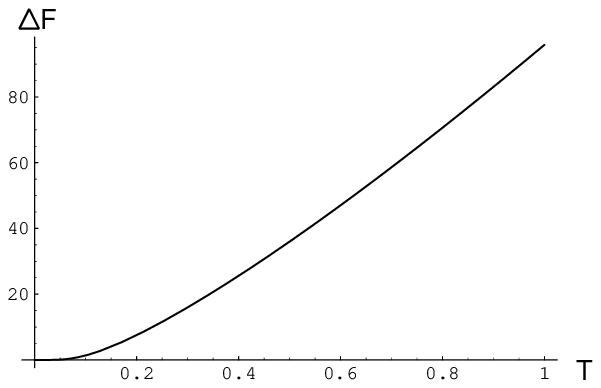} \qquad 
\includegraphics[scale=1.0]{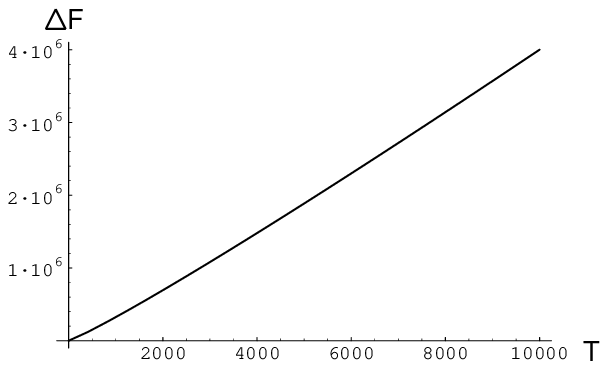}  \\
a) $\Delta F \equiv F_{\rm 2dim} - F_{\rm tri}$\,.  
\vspace*{0.7cm}\\
\includegraphics[scale=1.0]{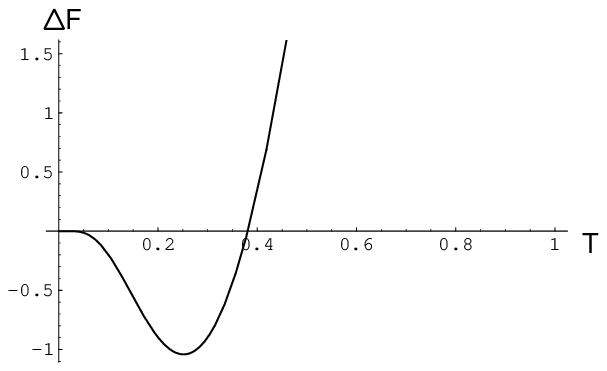} \qquad 
\includegraphics[scale=1.0]{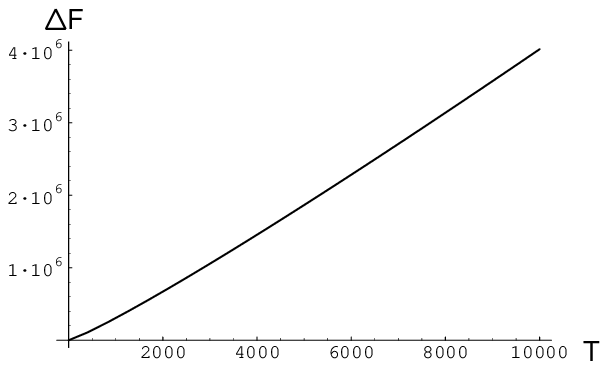}  \\
b) $\Delta F \equiv F_{\rm 2dim} - F_{\rm 3dim}$\,. 
\vspace*{0.7cm}\\
\includegraphics[scale=1.0]{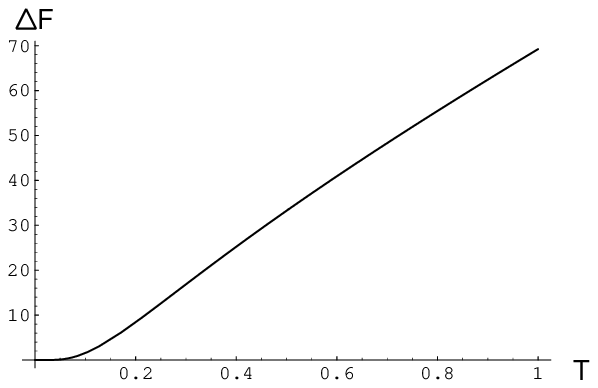} \qquad  
\includegraphics[scale=1.0]{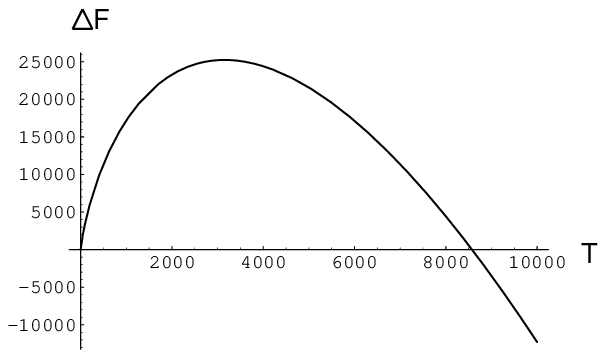}  \\ 
c) $\Delta F \equiv F_{\rm 3dim} - F_{\rm tri}$\,. \\
\caption{The numerical plots of the difference of free energies in the 
$N=3$ case. }
\label{n=3:fig}
 \end{center}
\end{figure}

\begin{figure}[htbp]
 \begin{center}
  \includegraphics{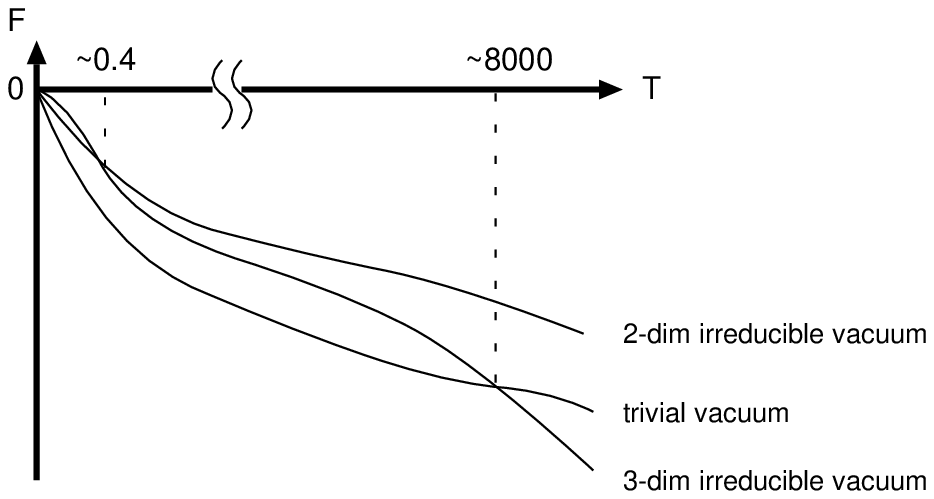} 
\caption{The sketch of free energies in the $N=3$ case.}
\label{sketch-n=3:fig}
 \end{center}
\end{figure}

We first consider the $N=3$ case.  In this case, there are three types
of vacua; i) trivial vacuum, ii) three-dimensional irreducible fuzzy
sphere, iii) two-dimensional irreducible fuzzy sphere. The third
vacuum iii) is represented by a direct sum of two-dimensional
irreducible representation of fuzzy sphere and one-dimensional trivial
vacuum.  The free energy for each vacuum is given by, respectively,
\begin{eqnarray}
\mbox{i)}~F_{\rm tri} = F_0[N=3]\,, \quad 
\mbox{ii)}~F_{\rm 3dim} = F[N=3]\,, \quad 
\mbox{iii)}~F_{\rm 2dim} = F[N=2] + F_0[N=1]\,,
\end{eqnarray}
where we assumed that there are no interactions between the trivial
vacuum and fuzzy sphere in the case iii)\,.  We can numerically
evaluate and compare three free energies.  The energy difference
between them are plotted in Fig.\,\ref{n=3:fig}.
>From Fig.\,\ref{n=3:fig}, we can read off the sketch for behaviors of
free energies, which is depicted in Fig.\,\ref{sketch-n=3:fig}.
Namely, once the system is coupled to the heat bath, the degeneracy of
vacuum at zero temperature is resolved and the trivial vacuum has the
lowest free energy. In the high temperature region $T\sim$ 8000, the
three-dimensional irreducible representation of fuzzy sphere is
favored rather than trivial vacuum.  Thus, the vacuum with the largest
irreducible representation of fuzzy sphere is most stable at
sufficiently high temperature. This result is the same way as in the
irreducible vacuum cases.

It should be noted that we can confirm the behavior of free energies
at low and high temperature by using the asymptotic forms.  In the low
temperature region, the free energies behave as
\begin{eqnarray}
F_{\rm tri} \sim - 54T \e^{-\frac{1}{6T}} \,, \quad 
F_{\rm 3dim} \sim - 6T \e^{-\frac{1}{6T}} \,,  \quad 
F_{\rm 2dim} \sim - 12T \e^{-\frac{1}{6T}} \,.  
\end{eqnarray}
On the other hand, the high temperature behaviors are described by 
\begin{eqnarray}
F_{\rm tri} \sim -81 T\ln T\,, \quad 
F_{\rm 3dim} \sim -89 T\ln T\,, \quad 
F_{\rm 2dim} \sim -48 T\ln T\,. 
\end{eqnarray}
These are compatible with the sketch in Fig.\,\ref{sketch-n=3:fig}, 
and thus we could analytically check the consistency of numerical results. 

\begin{figure}
 \begin{center}
  \includegraphics[scale=1.0]{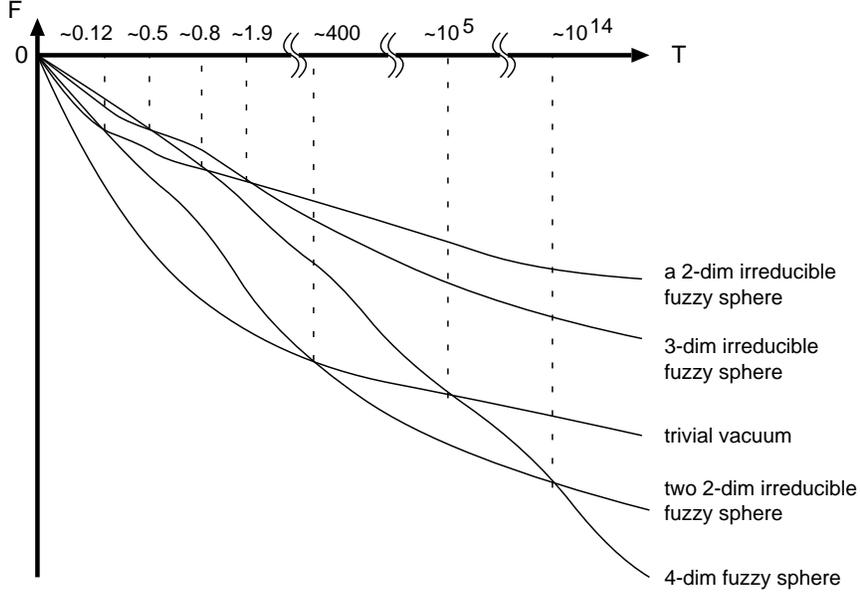}
\caption{The sketch of free energies in the $N=4$ case.}
\label{sketch-n=4:fig}
 \end{center}
\end{figure}

Next, we shall consider the $N=4$ case. Now we have five types of
vacua; i) trivial vacuum, ii) a couple of two-dimensional irreducible
fuzzy spheres, iii) four-dimensional irreducible fuzzy sphere, iv) a
two-dimensional fuzzy sphere, v) a three-dimensional fuzzy sphere.  It
is possible to investigate numerically the free energy behaviors as in
the $N=3$ case. Because the range of temperature is quite broad and
hence we need many graphs of numerical plots, we give only a sketch of
the free energies, which is shown in Fig.\,\ref{sketch-n=4:fig}.
It can be seen that the trivial vacuum is most stable at low
temperature. Then the vacuum with a couple of two-dimensional
irreducible fuzzy spheres tend to be preferred from $T\sim 400$.
Finally, the four-dimensional irreducible fuzzy sphere is favored at
sufficiently high temperature.  In the comparison with the $N=3$ case,
there is a middle state which is composed of two fuzzy spheres. Thus,
we may guess a transition process that $2\times 2$ fuzzy spheres are
formed in the trivial vacuum. After this creation process, they would
combine each other and form a four-dimensional irreducible
representation.

In the present case, it is also possible to confirm the low and high
temperature behaviors of free energies by using their asymptotic
forms.  In the low temperature region, the free energy in each of
vacua is expressed as, respectively,
\begin{eqnarray}
&& F_{\rm tri} \sim -96 T \e^{-\frac{1}{6T}}\,, \quad 
F_{2\times 2{\rm dim}} \sim -24 T \e^{-\frac{1}{6T}}\,, \quad 
F_{\rm 4dim} \sim - 6 T\e^{-\frac{1}{6T}}\,, \nn \\
&& F_{\rm 2dim} \sim - 30 T\e^{-\frac{1}{6T}}\,, \quad 
F_{\rm 3dim} \sim - 12 T\e^{-\frac{1}{6T}}\,.  
\end{eqnarray} 
The high temperature behaviors of free energies are 
\begin{eqnarray}
&& F_{\rm tri} \sim -144 T\ln T\,, \quad 
F_{2\times2{\rm dim}} \sim -156 T\ln T\,, \quad 
F_{\rm 4dim} \sim - 159 T\ln T\,, \nn \\
&& F_{\rm 2dim} \sim - 95 T\ln T\,, \quad 
F_{\rm 3dim} \sim - 98 T\ln T\,.  
\end{eqnarray} 
>From these asymptotic forms, we can estimate the initial and final
behaviors of free energies. As the result, the sketch in
Fig.\,\ref{sketch-n=4:fig} based on numerical studies is confirmed by
analytical evaluation.

As a matter of course, we may proceed with the numerical analysis of
free energies in the $N\geq 5$ case. We will not carry out here, but
it would be possible in principle for a fixed value of $N$ to check
that the $2\times 2$ fuzzy spheres are firstly formed as the
temperature grows, and then they are combined with each other or
enhance to larger irreducible representations.  Finally the maximal
size irreducible representation would be realized after some
transitions.  That is, we may argue that an $N$-dimensional
irreducible representation dominates a direct sum of small irreducible
representations of fuzzy spheres in the $U(N)$ pp-wave matrix model at
sufficiently high temperature.

\section{Conclusion and Discussion} 

We have discussed thermodynamics of fuzzy sphere in a matrix model on
a pp-wave background.  The exact free energy for an arbitrary matrix
size $N$ has been computed in the $\mu\rightarrow\infty$ limit.  This
is a generalization of the result obtained by Huang \cite{Huang} where
the $2\times 2$ matrix formulation \cite{SY3} was used. We have found
that the free energy in the fuzzy sphere vacuum is smaller than that
in the trivial vacuum above a critical temperature by carrying out a
numerical analysis of free energy with fixed matrix sizes.  This
result may imply that the fuzzy sphere vacuum is more stable than the
trivial vacuum at sufficiently high temperature, and would support
that the fluctuations around the trivial vacuum might condense to form
fuzzy sphere solutions above the critical temperature.

We have also seen that the critical temperature increases as the
matrix size $N$ becomes large.  In particular, our evaluation suggests
that the change in vacuum type would not happen in the $N\rightarrow
\infty$ limit, and it is consistent to the supergravity analysis.  In
addition, we have found the evidence that an $N$-dimensional
irreducible fuzzy sphere (i.e., the largest irreducible
representation) is more stable than the reducible one through some
numerical plots.

In conclusion, we may argue that a fuzzy sphere vacuum belonging to an
$N$-dimensional irreducible representation would dominate above a
certain temperature.  It is important to give an exact proof for this
argument. At any rate, we have to make an effort to clarify whether
the critical phenomenon such as a phase transition should exist or
not.

It is an interesting future problem to include higher order
contributions to our result by computing the energy shift.  It is also
interesting to study thermodynamics of other classical solutions (For
classical solutions in a pp-wave matrix model, for example, see
\cite{Bak,HS1,Park}).  On the other hand,the trivial vacuum $X=0$ is
related to the transverse spherical M5-brane \cite{TM5}. The
five-brane dynamics in the pp-wave matrix model is studied in the
recent work \cite{Furuuchi}.  It would be worthwhile to proceed to
study in this direction.  On the other hand, thermodynamic property of
type IIA string theory \cite{SY4,HS2} is studied in \cite{thermal-aa}.
It is also interesting to consider our results from the viewpoint of
type IIA string theory.  For the study in this direction, a matrix
string theory on a pp-wave, which is derived in \cite{SY4} by using
the method \cite{Sekino}, would be surely available.

\section*{Acknowledgments} 

One of us, K.Y., would like to thank J.~Nishimura, D.~Tomino and Y.~Kimura for
helpful discussions.  The work of H.S. was supported by Korea Research
Foundation (KRF) Grant KRF-2001-015-DP0082.

\end{document}